\documentstyle[psfrag,graphicx,aps,twocolumn]{revtex}
\def\be{\begin{eqnarray}}
\def\ee{\end{eqnarray}}

\def\la{\langle}
\def\ra{\rangle}

\def\roughly#1{\mathrel{\raise.3ex\hbox{$#1$\kern-.75em%
\lower1ex\hbox{$\sim$}}}}

\begin{document}

\renewcommand{\thefootnote}{\arabic{footnote}}
\setcounter{footnote}{0}

\vskip 0.4cm

\title{\LARGE\bf Free L\'{e}vy Matrices and Financial Correlations}

\author{
Zdzis\l{}aw Burda$^{a,b}$\thanks{E-mail: burda@physik.uni-bielefeld.de}, 
Jerzy Jurkiewicz$^{a}$\thanks{E-mail: jjurkiew@th.if.uj.edu.pl},
Maciej A. Nowak$^{a}$\thanks{E-mail: nowak@th.if.uj.edu.pl },\\
Gabor Papp$^{c,d}$\thanks{E-mail: pg@ludens.elte.hu}
and Ismail Zahed$^{c}$\thanks{E-mail: zahed@zahed.physics.sunysb.edu}}

\address{
$^a${\it M. Smoluchowski Institute of Physics,
Jagellonian University, Cracow, Poland} \\
$^b${\it Fakult\"at f\"ur Physik, Universit\"at Bielefeld
P.O.Box 100131, D-33501 Bielefeld, Germany} \\
$^c${\it Department of Physics and Astronomy,
SUNY-Stony-Brook, NY 11794  U.\,S.\,A.} \\
$^d${\it HAS Research Group for Theoretical Physics, 
E\"otv\"os University, Budapest, H-1518 Hungary}}

\date{\today}
\maketitle
\begin{abstract}
We consider a covariance matrix composed of asymmetric and free
random L\'{e}vy matrices. We use the results of free random variables
to derive an algebraic equation for the resolvent and solve it to
extract the spectral density. For an appropriate choice of 
asymmetry and L\'{e}vy index ($\alpha/2=3/4$) the free eigenvalue spectrum
is in remarkable agreement with the one obtained from the covariance 
matrix of the SP500 financial market. Our results are of interest to
a number of stochastic systems with power law noise.
\end{abstract}



\vskip .5cm
{\bf 1.\,\,\,}
The basic concept of independence of commuting random variables has
been generalized by Voiculescu to noncommuting ones such as random matrices,
using the powerful theory of free random variables~\cite{VOI}. The 
extension covers formally all stable distributions such as the L\'{e}vy 
ones. Free random matrices with gaussian fixed point, have been applied 
to many physical problems~\cite{GROSS,ZEE,QCD}.

Recently we have extended the concept of free random L\'{e}vy variables to
matrices~\cite{US00}, and suggested that the results may be relevant 
for addressing the issue of noise in stochastic systems with power
law distributions. The latters are encountered in physics 
(e.g. solar wind data), biophysics (e.g. heartbeat 
data) and finances (e.g. financial time series)~\cite{STAN,BOU}.

A quantitative way of addressing the issue of noise in gaussian
stochastic systems is through the use of covariance matrices. In
nongaussian systems, this concept is usually substituted by the
one of covariation~\cite{COVA} or tail covariance~\cite{BOU}. In this letter,
we will show that the concept of freeness as extended to matrices
\cite{US00}, allows for a simple analytical understanding of the
eigenvalue structure of the covariance matrix, where the underlying noise 
is inherently power law distributed. 

In section 2 we recall some results for free
random matrices, which will be useful in the further part
of the paper, with an emphasis on the Coulomb gas construction.
In section 3, we define the covariance for random asymmetric
L\'{e}vy matrices and derive a closed algebraic equation for the
resolvent in the large size limit and for fixed asymmetry. We solve
it to derive pertinent spectral distributions for L\'{e}vy covariances.
In section 4, we show that our results may be relevant to the 
understanding of financial spectral correlations, a point of recent interest
\cite{RAN}. Our conclusions are in section 5.

\vskip 1.5cm
{\bf 2.\,\,\,}
The resolvent for an $N\times N$ random matrix $M$ 
is generically given by
\be
G(z) =\frac 1N\,\,\left\la {\rm Tr}\left( \frac 1{z-M}\right)\right\ra\,\,.
\label{0}
\ee
In the following we consider the unitary ensemble of Hermitean matrices
(other ensembles can be studied similarly) in the large-$N$ limit.
The averaging is carried with the measure  $e^{-N\,{\rm Tr}\,V(M)}\,dM$. 
For standard random matrices the potential 
$V(M)$ is usually an even power series in $M$ bounded from below.
For free random L\'{e}vy matrices, the potential is not analytic in $M$.
It is known explicitly in two cases~\cite{US00}:
$V(M)={\rm ln}(M^2+b^2)$ (Cauchy) and
$V(M)=(1/M+ {\rm ln}\, M^2)$ (L\'{e}vy-Smirnov),
and implicitly as we now discuss.

The measure given above depends only on eigenvalues of the $M$ matrix. The
rotational degrees of freedom can be integrated out yieldieng  the Coulomb-gas
representation~\cite{DYSON} 
\be
\label{2}
\rho(\lambda_1,.\!.\!.,\lambda_N) \prod_i d\lambda_i = \prod_i d\lambda_i
e^{-NV(\lambda_i)} \prod_{i<j} (\lambda_i-\lambda_j)^2 \,\,,
\ee
where 
\be
\label{3}
V'(\lambda)=2\, \mbox{\rm Re }\, G(\lambda +i0)\,\,.
\ee
For free L\'{e}vy matrices $G(z)$ satisfies an algebraic equation
in the $N\to \infty$ limit ~\cite{US00} ($\alpha\neq 1$)
\be
\label{4}
b\,G^{\alpha}(z)-(z-a)\,G(z)+1=0\,\,,
\ee
in the upper half-plane, and follows by Cauchy reflection in the lower
half-plane. The parameter $b$ is related to the L\'{e}vy index $\alpha$, 
asymmetry $\beta$, and range or scale $\gamma$ \cite{US00}.
In particular, for symmetric distributions ($\beta=0$), it reads
\be
b = -\gamma\ e^{i\alpha\pi/2} \, .
\label{bgamma}
\ee 
The marginal case $\alpha=1$ is discussed in~\cite{US00}.
Equation (\ref{4}) can be solved analytically in many cases~\cite{US00}, and
numerically by starting at large $z$ and moving inward, using continuity 
and the asymptotic value of the physical branch
\be
G(z) \rightarrow 1/z + b/z^{1+\alpha}\,\,.
\ee
The spectral density is $\rho (\lambda ) = -{\rm Im}\,G(\lambda +i0)/\pi$.
For finite $N$ we expect corrections proportional to $1/N^p$ with some positive
$p$, depending on the ensemble. In the following we shall neglect these
corrections. 

\begin{figure}
\centerline{\includegraphics[width=0.48\textwidth]{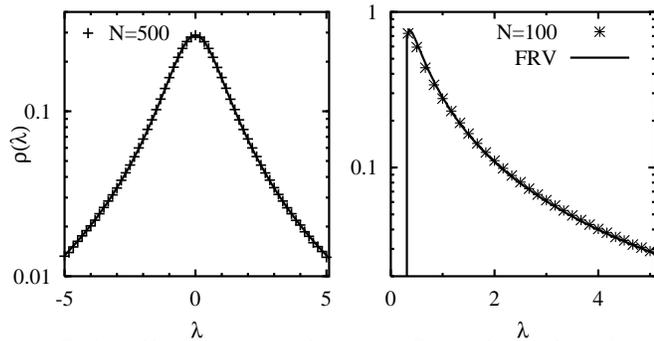}}
\caption{
Right: Metropolis Monte Carlo simulation of the Coulomb gas with
$\alpha=1/2$ and $\beta=1$
(crosses) compared to the FRV result (line). 
Left: the same for $\alpha=1$ and $\beta=0$
}
\label{fig1}
\end{figure}

To show that this is legitimate and that
(\ref{2}-\ref{4}) define correctly the free random L\'{e}vy
ensemble, we have numerically calculated the spectral density using
a Monte-Carlo simulation of the Coulomb gas in the Cauchy and L\'{evy}-Smirnov 
case for large, but finite $N$. In Fig.~\ref{fig1}a we show the results 
for the Cauchy case and in Fig.~\ref{fig1}b 
we show the results for the L\'{evy}-Smirnov case. For large $N$, the results 
asymptote $(b/\pi)/(\lambda^2+b^2)$ (Cauchy) and 
$\sqrt{4\lambda -1}/(2\pi\lambda^2)$ (L\'{evy}-Smirnov) shown in solid lines.
Other cases can be obtained by solving (\ref{3}-\ref{4}) to generate
the potential, and then Monte-Carlo sampling to generate any n-point
density. Note that the Cauchy and L\'{evy}-Smirnov
cases are directly accessible by random matrix sampling since the 
measure is explicitly known in these two cases (and only these two cases).

\vskip 1.5cm
{\bf 3.\,\,\,}
One way of quantifying the fluctuations in a 
stochastic system is through the use of
a covariance matrix ${\bf C}$,
\be
{\bf C}_{ij} =\frac 1{T^{2/\alpha}} \,\,\sum_{t=1}^T\,\,
M_{ti}\,M_{tj}\,\,,
\label{5}
\ee
where $M_{ti}$ is $T\times N$ (asymmetric).
The normalization in (\ref{5})
follows from the fact that the ``variance'' for L\'{e}vy matrices grows like
$T^{2/\alpha}$ where $\alpha=2$ is the expected diffusive limit (Brownian).
Free L\'{e}vy ensembles are super-diffusive for $0<\alpha<1$ and
sub-diffusive for $1<\alpha<2$ with $\alpha=1$ the critical divide.

The spectrum of ${\bf C}$ contains important information about the
character of the correlations. It follows from the resolvent $W(z)$
of ${\bf C}$. Since (\ref{5}) involves a product $MM^T$, the resolvent of 
${\bf C}$ can be related to the resolvent of $M$  using Voiculescu's
powerful machinery of $R$ and $S$ transforms for free random 
variables~\cite{VOI}. 
Following the procedure described in \cite{QCD},
 in our case we find that
$W(z)$ satisfies the transcendental equation
\be
\gamma^{2/\alpha} W(z\gamma^{2/\alpha})= 
W_{\gamma=1}(z) \equiv
\frac{1+w(z)}{z} \,,
\label{Ww}
\ee
where $w(z)$ satisfies the multi-valued equation
\be
- e^{i \frac{2\pi}{\alpha}} \cdot w^{2/\alpha} \cdot z =(1+w)(w+m)\,\,,
\label{wp}
\ee
for fixed asymmetry $m=T/N$, and L\'evy asymmetry parameter $\beta=0$.
The range of the L\'evy distribution enters the formula only
through the prefactor of $W(z)$, and the normalization
of $z$. Thus, using the scaling on the left-hand side of (\ref{Ww}),
one can easily rescale the $\gamma=1$ result to any other value of
$\gamma$. The phase factor on the left hand side of (\ref{wp})
can be deduced from (\ref{bgamma}).

The distribution of eigenvalues of the covariance matrix follows from
the discontinuity of $w(z)$,
\be
  \rho(\lambda) = \frac1{\pi\lambda} \mbox{Im}\, w(\lambda +i0) \,.
\ee
The density of eigenvalues of index $\alpha$ and range $\gamma$ 
satisfies the self-affine property 
$  \rho_{\gamma}(\lambda) = 
	\rho(\lambda/ \gamma^{2/\alpha})/\gamma^{2/\alpha} $.
Its tail-behavior is given by
\be
  \rho(\lambda) \approx \frac1{\pi} \frac{m^{\alpha/2}\sin{\alpha	
   \frac{\pi}2}} {\lambda^{1+ \alpha /2}}\,\,\, ,
\label{powertail2}
\ee
for $\lambda \gg \lambda_{\rm min} > 0$ and symmetric ($\beta=0$) entries. 
The distribution is unimodal
and unbounded from above. Since the covariance is a square, the
tail distribution is characterized by an index $\alpha/2$ since
the entries $M_{ti}$, are L\'{e}vy distributed with index $\alpha$. 

\begin{figure}
\centerline{\includegraphics[width=0.48\textwidth]{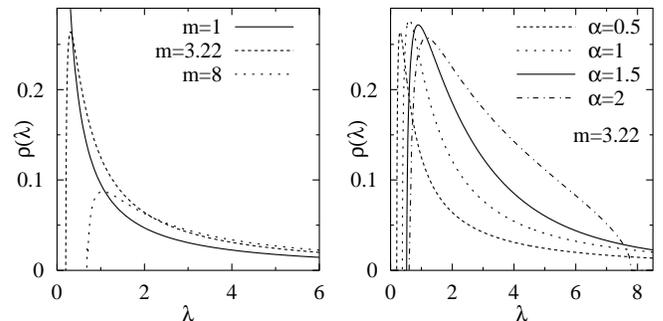}}
\caption{
Left: Spectral density of FRV with  $\alpha=1/2$ and different 
asymmetry parameters $m$. Right: spectral density for 
several indices $\alpha$, at $m=3.22$. 
}
\label{fig2}
\end{figure}
For a symmetric Cauchy distribution ($\alpha=1$), 
the above analysis yields for the resolvent
\be
W_{\gamma=1}(z) = \frac 1z \left(\! 1\!-
\frac{(m\!+\!1)\mp 
\sqrt{(m\!-\!1)^2\!-\!4mz}}{2(1+z)}\,\right)\,\,,
\label{CAU1}
\ee
from which the density of eigenvalues follows
\be
\rho (\lambda ) = \frac 1\pi \frac{\sqrt{m}}{\lambda (\lambda + 1)}
\,{(\lambda -\lambda_{\rm min})}^{1/2}\,\,,
\ee
with $\lambda_{\rm min} = (m-1)^2/4m$. The spectrum starts at
the minimum eigenvalue $\lambda_{\rm min}$ and is unbounded from above. 
In general the minimal eigenvalue is determined by
\be
  \lambda_{\rm min} &=& -\frac{(1+w_*) (m+w_*)}{(-w_*)^{2/\alpha}}
\ee
where $w_*$ is the negative solution of the second order equation
\be
  (\alpha-1) w_*^2-(1-\frac{\alpha}2)(m+1) w_* - m =0 \,.
\ee
Some typical spectra are shown in Fig.~\ref{fig2}a. 
We now illustrate the relevance of these distributions
for the covariance spectrum in a financial market,
a point of current interest~\cite{RAN}.

\vskip 1cm
{\bf 4.\,\,\,}
Recently, it was pointed out that financial covariances are
permeated by Gaussian noise in the low-lying eigenvalue region 
with consequences on risk assessment~\cite{RAN}. Here we show that
the financial covariance spectrum is throughout 
permeated by free L\'{e}vy noise.
For that consider the empirical matrix of relative returns
\be
\label{M1} 
C_{ij} &=& \frac1T \sum_t M_{ti} M_{tj} \qquad\mbox{with} \\
M_{ti} &=& \left( x_{it+1}-x_{it}\right)/x_{i0} - \langle M\rangle\,\,,
\nonumber
\ee
where $x_{it}$ are the raw returns made of the price $x_{it}$ 
of stock i at time t. Several other definitions are also posiible, 
see~\cite{USII}. The returns are normalized to the initial
price $x_{i0}$ to insure scale invariance. They carry zero mean
after subtracting the average relative return
$\langle M\rangle$. For the raw
prices we will use the daily quotations of $N=406$ stocks 
from the SP500 market over the period of $T+1=1309$ days 
from 01.01.1991 till 06.03.1996 (ignoring dividends). For these
data the matrix asymmetry is $m=1308/406\approx 3.22$.

\begin{figure}
\centerline{\includegraphics[width=0.48\textwidth]{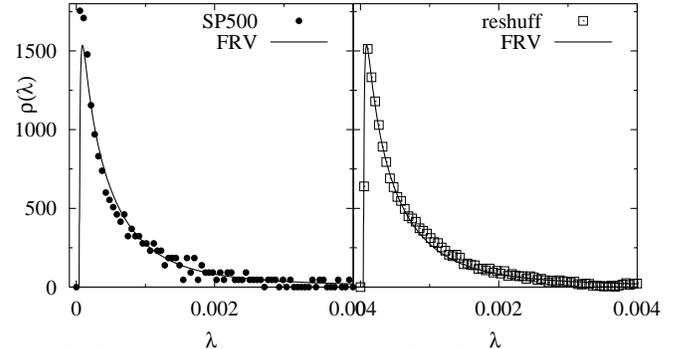}}
\caption{ 
The spectrum of the historically ordered data
(left) and the reshuffled data averaged over 20 reshufflings
(right) for SP500 compared
to the FRV result with $\alpha=1.5$ and $m=2$.
}
\label{fig3}
\end{figure}

In Fig.~\ref{fig3} we compare the analytical results following from the free 
L\'{e}vy covariance to the ones following from the SP500 market setting
the scale of the free covariance to match the one of the market. 
Fig.~\ref{fig3} left 
shows the distribution of eigenvalues for free L\'{e}vy matrices
with index $\alpha/2=3/4$ and asymmetry $m=2$, versus the raw SP500 data.
Fig.~\ref{fig3} right shows
the same distribution of eigenvalues for free L\'{e}vy matrices  
versus  the reshuffled SP500 data.
The reshuffled data follow from a random
reshuffling of the price series for a fixed stock, destroying all
inter-stock correlations. Our optimal fit preserves the index of
the cumulative distribution, albeit for a smaller asymmetry ($m=2 <3.22$).
It is remarkable that our free L\'{e}vy fit to the spectrum of the covariance
suggests that the time series of returns is power law distributed with index
$\alpha=3/2$ which is consistent with detailed studies of the SP500
financial time series~\cite{BOU}.

Finally, we would like to mention that there exists an alternative
construction for random matrices with spectral L\'{e}vy disorder, 
as discussed in~\cite{BOUCIZ}, where the matrix elements are 
populated from  independent , one-dimensional stable distributions.
As a result, these Random L\'{e}vy Matrix (RLM) ensembles
are not rotational invariant, contrary to 
the Free L\'{e}vy Matrix (FLM) ensembles discussed here. 
We note that RLM ensembles exhibit interesting correlations
between the eigenvectors, which are absent (or rather trivial) in the case
of the rotation invariant ensembles such as the GUE, GOE, GSpE or FLM ones.   
For detailed studies of covariances based on RLM ensembles we refer to~\cite{USII}.

\vskip 1cm
{\bf 5.\,\,\,}
We have shown that the resolvent of covariances constructed 
from free random L\'{e}vy matrices, obeys an algebraic equation
for all $1\leq \alpha\leq 2$. The ensuing distribution of eigenvalues was
constructed for a range of $\alpha$'s. We have argued that this
distribution may be relevant to a large number of problems where
power law fluctuations are expected, an example being the SP500
market. Indeed, we have shown that the spectral density of the
SP500 financial covariance is in overall agreement with a free
L\'{e}vy distribution of index $\alpha/2=3/4$. Free L\'{e}vy noise 
may be dominant in financial covariances, a point of relevance
to the central issue of assessing risk in finances. 

\vskip 1cm
{\bf Acknowledgments:}
\vskip .5cm
This work was supported in part by the US DOE grant DE-FG02-88ER40388,
by the Hungarian FKFP grant 220/2000 and the EC IHP grant HPRN-CT-1999-00161.


\begin{thebibliography}{99}
\vspace*{-1cm}

\bibitem{VOI}
D.V. Voiculescu, {\it Invent. Math.} {\bf 104} (1991) 201;
D.V. Voiculescu, K.J. Dykema and A. Nica, ``Free Random Variables'',
Am. Math. Soc., Providence, RI (1992);
H. Bercovici and D. Voiculescu, {\it Ind. Univ. Math. J.}{\bf 42} (1993) 733.

    
\bibitem{GROSS}
R. Gopakumar and D. Gross, {\tt e-print hep-th/9411021}


\bibitem{ZEE}
A. Zee, {\it Nucl. Phys.} {\bf B474} )1996) 726;
J. Feinberg and A. Zee, {\it Nucl. Phys.} {\bf B501} (1997) 643; 
{\it ibid.} {\bf B504} (1997) 579.


\bibitem{QCD}
R.A. Janik, M.A. Nowak, G. Papp and I. Zahed, 
{\it Acta Phys. Pol.} {\bf B28} (1997) 2949; {\tt e-print hep-th/9710103};
R.A. Janik, M.A. Nowak, G. Papp, J. Wambach and I. Zahed,
{\it Phys. Rev.} {\bf  E55} (1997) 4100;
R.A. Janik, M.A. Nowak, G. Papp and I. Zahed, 
{\it Nucl. Phys.} {\bf B501} (1997) 603.

\bibitem{US00}
Z. Burda, R. Janik, J. Jurkiewicz, M.A. Nowak, G. Papp and I. Zahed, 
{\it Free random L\`{e}vy Matrices}, {\tt e-print cond-mat/0011451}.

\bibitem{STAN}
R.~Mantegna and H.~Stanley, 
{\it An Introduction to Econophysics}, Cambridge Univ. (2000). 

\bibitem{BOU}
J. Bouchaud and M. Potters, {\it Theory of Financial Risks}, Cambridge Univ.
(2000).
 

\bibitem{COVA}
G. Samorodnitsky and M. Taqqu, {\it Stable Non-Gaussian 
Random Processes}, Chapman and Hall (1994).


\bibitem{RAN}
L.~Laloux, P.~Cizeau, J.~Bouchaud and M.~Potters,
{\it Phys. Rev. Lett.} {\bf 83}  (1999) 1467, {\tt e-print cond-mat/9810255};
V.~Plerou, P.~Gopikrishnan, B.~Rosenow, L.~Nunes~Amaral, and H.~Stanley,
{\it Phys. Rev. Lett.} {\bf 83} (1999) 1471, {\tt e-print cond-mat/9902283}.



\bibitem{DYSON}
for  reviews on random matrix models and techniques, see e.g. 
M.L. Mehta, {\it ``Random Matrices''}, Academic Press, New York (1991);
T. Guhr, A. M\"{u}ller-Gr\"{o}ling and H.A. Weidenm\"{u}ller
,{\it Phys. Rep.}{\bf 299} (1998) 189;
P.  Di Francesco, P. Ginsparg and J. Zinn-Justin, {\it Phys. Rept.}
{\bf 254} (1995) 1, and references therein.

\bibitem{BOUCIZ}
P. Cizeau and J.-P. Bouchaud, {\it Phys. Rev.}
{\bf E50} (1994) 1810. 

\bibitem{USII}
Z. Burda, J. Jurkiewicz, M.A. Nowak, G. Papp and I. Zahed,
{\it L\'{e}vy Matrices and Financial Covariances}, {\tt e-print cond-mat/0103108}.



\end{thebibliography}
\end{document}